\documentclass[%
 reprint,
 superscriptaddress,
 preprintnumbers,
 nofootinbib,
 amsmath,amssymb,
 aps,
 prl,
]{revtex4-1}
\usepackage{psfrag,slashed,cancel,array,graphicx,todonotes,hyperref}
\usepackage{subfigure}
\usepackage[labelfont=bf,labelsep=quad,justification=justified]{caption}
\usepackage[utf8]{inputenc}
\usepackage{mathtools}
\usepackage{mathrsfs}
\usepackage{multirow}
\usepackage{braket}
\usepackage{booktabs} 
\usepackage[normalem]{ulem}
\hypersetup{pdftitle={},pdfcreator={},linkcolor=[rgb]{0.15,0.35,0.75},colorlinks=true,citecolor=[rgb]{0.675,0,0.2},urlcolor=[rgb]{0.15,0.35,0.65}}

\def\beq{\begin{equation}}
\def\eeq{\end{equation}}
\def\bsp#1\esp{\begin{split}#1\end{split}}

\newcommand{\refcite}[1]{ref.~\cite{#1}}
\newcommand{\refscite}[1]{refs.~\cite{#1}}
\newcommand{\Eq}[1]{Eq.~\eqref{eq:#1}}

\newcommand{\eq}[1]{eq.~\eqref{eq:#1}}

\newcommand{\fig}[1]{figure~\ref{fig:#1}}

\newcommand{\cJ}{\mathcal{J}}

\newcommand{\cN}{\mathcal{N}}
\newcommand{\cO}{\mathcal{O}}
\newcommand{\GammaC}{\Gamma_{\rm cusp}}
\newcommand{\nn}{\nonumber}

\newcommand{\born}{ \hat \sigma_0}
\newcommand{\df}{\mathrm{d}}
\newcommand{\bn}{{\bar n}}
\newcommand{\bq}{{\bar q}}
\newcommand{\SCETII}{SCET$_{\text{II}}$ }
\newcommand{\as}{\alpha_s}
\newcommand{\bt}{{\vec b}_T}
\newcommand{\qt}{{\vec q}_T}
\newcommand{\blue}[1]{{\color{blue}{#1}}}
\newcommand{\gammathr}{\gamma_{\rm th}}
\newcommand{\SYM}{$\mathcal{N}=4$~sYM\,}
\newcommand{\img}{\mathrm{i}}

\AtBeginDocument{
\heavyrulewidth=.08em
\lightrulewidth=.05em
\cmidrulewidth=.03em
\belowrulesep=.65ex
\belowbottomsep=0pt
\aboverulesep=.4ex
\abovetopsep=0pt
\cmidrulesep=\doublerulesep
\cmidrulekern=.5em
\defaultaddspace=.5em
}

\def\cO{\mathcal{O}}
\def\cN{\mathcal{N}}

\begin{document}

\preprint{BONN-TH-2022-11, SLAC-PUB-17675}

\title{The Four-Loop Rapidity Anomalous Dimension and \\Event Shapes to Fourth Logarithmic Order}

\author{Claude Duhr}
\email{cduhr@uni-bonn.de}
\affiliation{Bethe Center for Theoretical Physics, Universit\"at Bonn, D-53115, Germany.}
\author{Bernhard Mistlberger}
\email{bernhard.mistlberger@gmail.com}
\affiliation{SLAC National Accelerator Laboratory, Stanford University, Stanford, CA 94039, USA.}
\author{Gherardo Vita}
\email{gherardo@slac.stanford.edu}
\affiliation{SLAC National Accelerator Laboratory, Stanford University, Stanford, CA 94039, USA.}

\begin{abstract}
We obtain the quark and gluon rapidity anomalous dimension to fourth order in QCD. 
We calculate the N$^3$LO rapidity anomalous dimensions to higher order in the dimensional regulator and make use of the soft/rapidity anomalous dimension correspondence in conjunction with the recent determination of the N$^4$LO threshold anomalous dimensions to achieve our result.
We show that the results for the quark and gluon rapidity anomalous dimensions at four loops are related by generalized Casimir scaling.
Using the N$^4$LO rapidity anomalous dimension, we perform the resummation of the Energy-Energy Correlation in the back-to-back limit at N$^4$LL, achieving for the first time the resummation of an event shape at this logarithmic order. We present numerical results and observe a reduction of perturbative uncertainties on the resummed cross section to below 1\%.
\end{abstract}

\maketitle

\section{Introduction}
Over the last decade we have entered a new era in QCD phenomenology, where we can perform high-precision computations for key observables, ranging from non-perturbative determinations of form factors from lattice QCD to high-order perturbative computations for high-energy collider processes. In many cases, perturbative computations develop large logarithms order by order in perturbation theory which may spoil the convergence of the perturbative series in the coupling constant. 
In such a scenario, these logarithms need to be resummed to all orders by a renormalization group equation (RGE) governed by certain anomalous dimensions. 
The most prominent examples of QCD anomalous dimensions are the QCD beta function and the cusp anomalous dimension, which control the running of the strong coupling and the structure of infrared singularities. 
The knowledge of QCD anomalous dimensions is therefore not only important to obtain precise phenomenological predictions, but they are also a unique window into the all-order perturbative structure of the strong interactions.
In the remainder of this Letter we compute the four-loop corrections of such an anomalous dimension, and we apply it to resum for the first time large logarithmic corrections to an event shape and a transverse momentum dependent (TMD) observable to fourth logarithmic order.

Examples of the class of collider observables characterized by the presence of large rapidity logarithms are transverse momentum distributions at hadron colliders. Their all-order resummation is dictated by the so-called rapidity anomalous dimensions~\cite{Chiu:2012ir}, which is closely related to the Collins-Soper kernel \cite{Collins:1981uk,Collins:1981va,Vladimirov:2020umg}.
Let's consider for concreteness the case of Drell-Yan like processes. The leading power factorization theorem for transverse momentum distributions, which encodes the all-order behavior in the coupling in the $q_T \to 0$ limit, can be written as \cite{Collins:1981uk,Collins:1981va,Collins:1984kg,Catani:2000vq,deFlorian:2001zd,Catani:2010pd,Becher:2010tm,Becher:2011xn,Becher:2012yn,GarciaEchevarria:2011rb,Echevarria:2012js, Echevarria:2014rua,Chiu:2012ir,Li:2016axz,Vladimirov:2020umg,Billis:2019vxg,Ebert:2020yqt,Luo:2019szz}

\begin{align} \label{eq:TMD_factorization}
& \frac{\df\sigma}{\df Q^2 \df Y \df^2 \qt} 
 = \sigma_0 \sum_{a,b} H_{ab}(Q^2,\mu) \int\!\frac{\df^2\bt}{(2\pi)^2} e^{\img\,\qt \cdot \bt}\nn\\
   &\times \tilde B_a\Bigl(x_1^B, b_T, \mu, \nu\Bigr)
   \,\tilde B_b\Bigl(x_2^B, b_T, \mu, \nu \Bigr)\, S_q(b_T, \mu, \nu)
\end{align}
The logarithmic dependence on the transverse momentum can be resummed by deriving RGEs for the objects appearing in the factorization theorem. For example, the soft function in \eq{TMD_factorization} obeys the following RGEs 
\begin{align}
	\mu \frac{\df}{\df \mu} \ln S_i(\bt, \mu, \nu) &= 4 \GammaC^i[\as(\mu)] \ln \mu/\nu + \gammathr^i[\as] \\
	\nu \frac{\df}{\df \nu} \ln S_i(\bt, \mu, \nu) &= -4 \int_{b_0/b_T}^\mu \frac{\df \mu^\prime}{\mu^\prime}\GammaC^i[\as(\mu^\prime)] + \gamma_r^i[\as]\,.\nn
\end{align}
where $i\in\{q,g\}$ labels the parton species and $\GammaC$ is the cusp anomalous dimension \cite{Korchemsky:1987wg,Bern:2005iz,Henn:2019swt,vonManteuffel:2020vjv}, $\gammathr^i$ is called the threshold anomalous dimension, and $\gamma_r^i$ is the rapidity anomalous dimension~\cite{Chiu:2012ir}. 
The rapidity anomalous dimension respects itself an RGE,
\beq\label{eq:gammarRGE}
\mu \frac{\df}{\df \mu} \gamma_r^i(b_T,\mu) = -4  \GammaC^i[\as(\mu)]\,,
\eeq
with the solution given by
\beq
	\gamma_r^i(b_T,\mu) = -4\int_{\mu_0}^\mu \frac{\df \mu^\prime}{\mu^\prime} \GammaC^i[\as(\mu')] +\gamma_r^i(\mu_0, b_T)\,,
\eeq
where $\mu_0$ is an arbitrary scale that marks the starting point of the RGE in \eq{gammarRGE}. 
Choosing $\mu_0 = b_0/b_T$ sets the logarithms in $\gamma_r^q(\mu_0, b_T)$ to zero and allows to express the boundary of the RGE as 
\beq 
	\gamma_r^i(\mu_0 = b_0/b_T, b_T) \equiv   \gamma_r^i[\as(b_0/b_T)]\,.
\eeq
The rapidity anomalous dimension boundary $\gamma_r^i[\as(b_0/b_T)]$ has been calculated to 3 loops in \refscite{Li:2016axz,Li:2016ctv}, and its calculation to 4 loops is one of the main results of this Letter. 

\section{Rapidity Anomalous Dimension at N$^4$LO}\label{sec:rapidity_anom_dim}

In \refscite{Vladimirov:2016dll,Vladimirov:2017ksc} an identity relating the threshold and rapidity anomalous dimensions using a conformal mapping of matrix elements of Wilson lines was proposed. 
In $d=4-2\epsilon$ dimensions the QCD beta function reads
\beq
	\beta[\as, \epsilon] = -2 \as\left[ \epsilon + \frac{\as}{4\pi} \beta_0 + \left(\frac{\as}{4\pi}\right)^2 \beta_1 + \dots \right] \,.
\eeq
Massless QCD is conformal up to the running of the strong coupling.
Consequently, there exists a critical point $\epsilon^*$ such that $\beta[\as, \epsilon^*] =0$ and QCD can be rendered conformal at this point.
Following \refcite{Vladimirov:2016dll,Vladimirov:2017ksc}, at this critical point the sum of the rapidity and threshold anomalous dimension vanishes
\beq \label{eq:rap_th_correspondence}
	\gamma_r^i[\as,\epsilon^*] + \gamma_{\textrm{th}}^i[\as,\epsilon^*] = 0\,.
\eeq
Since $\epsilon^* \sim \cO(\as)$, \eq{rap_th_correspondence} allows to obtain the standard rapidity anomalous dimension $\gamma_r^i[\as,0]$ in $d=4$ at $\cO(\alpha_s^n)$ from the $d=4$ threshold anomalous dimension at $\cO(\alpha_s^n)$ and the $d$-dimensional rapidity anomalous dimension $\gamma_r^i[\as,\epsilon]$ at $\cO(\alpha_s^{n-1})$.
In this Letter we have calculated the $d$-dimensional rapidity anomalous dimension to 3 loops extending the computation of the transverse momentum dependent soft function
\cite{Chiu:2012ir} of refs ~\cite{Li:2016ctv,Ebert:2020lxs,Ebert:2020yqt}
to higher orders in the dimensional regulator. Together with the 4-loop threshold anomalous dimension~\cite{Duhr:2022xxx,Das:2019btv}, we use it to extract the rapidity anomalous dimension for a parton species $i$ in representation $R$ to N$^4$LO. Identifying the coefficients of the perturbative expansion of the rapidity anomalous dimension as
\begin{align}
	\gamma_{r}^i [\as(\mu)] = \sum_{n} \left(\frac{\as(\mu)}{4\pi}\right)^n \gamma_{r,n}^i\,, 
\end{align}
the 4-loop coefficient reads
{\footnotesize
\begin{align}\label{eq:gammar4}
	\gamma_{r,4}^i &= \blue{C_A^3 C_R}\left(-\frac{21164}{9} \zeta_3^2-\frac{26104}{9} \zeta_2 \zeta_3+\frac{4228}{3} \zeta_4 \zeta_3+\frac{2752}{3} \zeta_2 \zeta_5
	\right.\nn\\&\qquad+\left.
		\frac{1201744 \zeta_3}{81}+\frac{778166 \zeta_2}{243}+\frac{8288 \zeta_4}{9}-\frac{181924 \zeta_5}{27}
	\right.\nn\\&\qquad-\left.
		\frac{63580 \zeta_6}{27}+\frac{11071 \zeta_7}{3}-\frac{28290079}{2187} -\frac{b^4_{q,C^4_{AF}}}{6}\right)
      	\nn\\&+ 
 		\blue{C_A C_R n_f^2}\left(\frac{224}{9} \zeta_3 \zeta_2+\frac{6752 \zeta_2}{243}-\frac{22256 \zeta_3}{81}+\frac{160 \zeta_4}{9}+\frac{1472 \zeta_5}{9}
   	\right.\nn\\&\qquad-\left.   
		\frac{898033}{2916}\right)	+	\blue{C_R n_f^3}\left(\frac{160 \zeta_3}{9}-\frac{16 \zeta_4}{9}+\frac{10432}{2187}\right)
	\nn\\&+
		\blue{C_R C_A^2 n_f}\left(-\frac{8584}{9} \zeta_3^2+\frac{2080}{3} \zeta_2 \zeta_3-\frac{247652 \zeta_3}{81}-\frac{182134 \zeta_2}{243}
	\right.\nn\\&\qquad+\left.
		\frac{43624 \zeta_4}{27}-\frac{17936 \zeta_5}{27}+\frac{1582 \zeta_6}{27}+\frac{10761379}{2916}	
	\right.\nn\\&\qquad-\left.
		\frac{b^4_{q,C^4_{FF}}}{12}-2 b^4_{q,n_f C_F^2 C_A}-b^4_{q,n_f C_F^3}\right)
	\nn\\&+
		\blue{C_RC_F n_f^2}\left(\frac{6928 \zeta_3}{27}+\frac{160 \zeta_4}{3}+32 \zeta_5-\frac{110059}{243}\right) 
	\nn
		\end{align}
		\begin{align}
		\phantom{	\gamma_{r,4}^i }
	\nn&+ 
		\blue{\frac{C^4_{AR}}{d_R}}\left(\frac{6688 \zeta_3^2}{3}+3584 \zeta_2 \zeta_3+736 \zeta_4 \zeta_3 +\frac{15616 \zeta_3}{9}-\frac{224 \zeta_4}{3}
	\right.\nn\\&\qquad+\left.
		\frac{4352 \zeta_2}{3}-2048 \zeta_2 \zeta_5+\frac{3680 \zeta_5}{9}-\frac{6952 \zeta_6}{9}-6968 \zeta_7
	\right.\nn\\&\qquad-\left.
		384 +4 b_{4,d4AF}\right)
	\nn\\&+ 
		\blue{\frac{C^4_{FR}}{d_R}n_f}\left(-\frac{2432}{3}\zeta_3^2-256 \zeta_2 \zeta_3+\frac{10624 \zeta_3}{9}-\frac{9088 \zeta_2}{3}
	\right.\nn\\&\qquad\left.
		+\frac{1600 \zeta_4}{3}+\frac{43520 \zeta_5}{9}-\frac{2368 \zeta_6}{9}+768+4 b^4_{q,C^4_{FF}}\right)
   	\nn\\&+ 
   		\blue{C_A C_F C_R n_f}\left(4 b_{4,n_f C_F^2 C_A}+\frac{6800 \zeta_3^2}{3}-\frac{8864}{9} \zeta_2\zeta_3-\frac{1892 \zeta_3}{9}
	\right.\nn\\&\qquad+\left.   
		\frac{5122 \zeta_2}{27}-\frac{122216 \zeta_4}{27}+\frac{21904 \zeta_5}{9}-1436 \zeta_6+\frac{2149049}{486}\right)
      	\nn\\&+ 
		\blue{C_F^2 C_R n_f}\left(4 b^4_{q,n_f C_F^3}-736 \zeta_3^2+\frac{1024}{3} \zeta_2 \zeta_3+\frac{2240 \zeta_3}{9}-648\zeta_2
	\right.\nn\\&\qquad+\left.   
		668 \zeta_4-\frac{7744 \zeta_5}{3}+\frac{29336 \zeta_6}{9}-\frac{27949}{54}\right) \,.		
\end{align}}%
We include the quark and gluon rapidity anomalous dimension in analytic form as electronically readable files together with the arXiv submission of this Letter.
Note that the rapidity anomalous dimension is expressed in terms of 4 coefficients
which currently have only been determined numerically in table 1 of ref.~\cite{Das:2019btv}.
We print their values here for convenience:
\begin{eqnarray}\label{eq:bdef}
b^4_{q,\,n_fC_F^2C_A}&=&-455.247 \pm 0.005\,,\\
b^4_{q,\,C^4_{AF}}&=&-998.0 \pm 0.2\,,\nonumber\\
b^4_{q,\,C^4_{FF}}&=&-143.6 \pm 0.2\,, \nonumber\\
b^4_{q,\,n_fC_F^3}&=&80.780 \pm 0.005\,. \nonumber
\end{eqnarray}
The high numerical precision allows us to obtain a very precise determination of the rapidity anomalous dimension for adjoint and fundamental Wilson lines:
{\small
\begin{align}
\label{eq:gammarnumerics}
	\gamma_r^q(n_f\! =\! 3) &=  0.39912 \as^2 + 0.52516 \as^3 + \big(0.60100 \pm 5\cdot 10^{-5}\big) \as^4\,, \nn\\
	\gamma_r^q(n_f\! =\! 4) &=  0.46924 \as^2 + 0.61648 \as^3 + \big(0.61362 \pm 5\cdot 10^{-5}\big) \as^4\,,\nn\\ 
	\gamma_r^q(n_f\! =\! 5) &=  0.53929 \as^2 + 0.68947 \as^3 + \big(0.53595 \pm 5\cdot 10^{-5}\big) \as^4\,,\nn\\[1mm]
	\gamma_r^g(n_f\! =\! 3) &=  0.89819 \as^2 + 1.18162 \as^3 + \big(1.5531 \pm 5\cdot 10^{-4}\big) \as^4\,, \nn\\
	\gamma_r^g(n_f\! =\! 4) &=  1.05580 \as^2 + 1.38708 \as^3 + \big(1.5884 \pm 5\cdot 10^{-4}\big) \as^4\,,\nn\\ 
	\gamma_r^g(n_f\! =\! 5) &=  1.21341 \as^2 + 1.55130 \as^3 + \big(1.4206 \pm 5\cdot 10^{-4}\big) \as^4\,,
\end{align}
}%
where the uncertainty on the N$^4$LO coefficients is estimated by propagating the uncertainties in \eq{bdef}. 
Since they affect the fourth/fifth significant digit of the N$^4$LO correction, we will treat them as negligible for the rest of this Letter.
Eqs.\eqref{eq:gammar4} and \eqref{eq:gammarnumerics} are some of the main results of this Letter.

\begin{figure}
 \centering
  \includegraphics[width=0.48\textwidth]{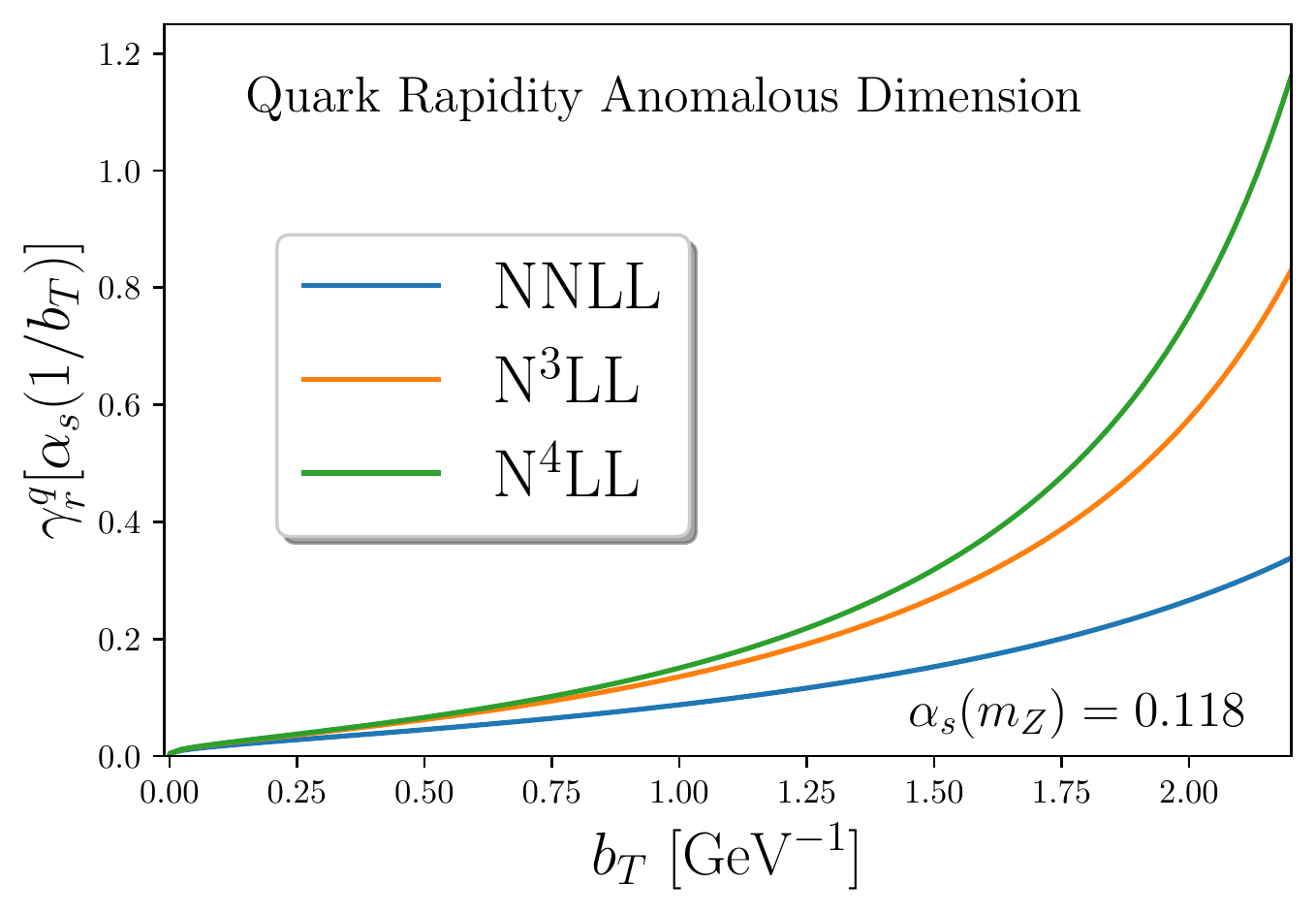}
 \caption{Boundary term of the rapidity quark anomalous dimension as a function of $b_T$ through four loops. The $b_T$ dependence enters only through the coupling constant. The diverging behavior  at large $b_T$ is due to approaching the Landau pole. For recent work on extracting the anomalous dimension non-perturbatively at large $b_T$ see \refscite{Ebert:2018gzl,Ebert:2019okf,Vladimirov:2020ofp,Scimemi:2019cmh,Bacchetta:2019sam,Ebert:2019tvc,Shanahan:2020zxr,LatticeParton:2020uhz,Schlemmer:2021aij,Li:2021wvl,Ebert:2022fmh}.}
 \label{fig:gammarplot}
\end{figure}

Let us emphasize that our result for $\gamma^i_{r,4}$ is essentially identical, analytically, for quarks and gluons, and only depends on the color representation through the quadratic and quartic Casimir operators 
\begin{align}
C_R&\, = \frac{1}{d_R}\text{tr}(T_R^aT_R^a)\,,\\
\nonumber C^4_{R'R}&\, = \frac{1}{(4!)^2} \text{tr}\left(T_{R'}^{\{a_1}\cdots T_{R'}^{a_4\}}\right)\text{tr}\left(T_{R}^{\{a_1}\cdots T_{R}^{a_4\}}\right)\,,
\end{align}
where $R'\in\{F,A\}$, $T_R^a$ are the generators of the representation $R$ and $d_R$ is the dimension of the color representation. 
This property is referred to as \emph{generalized Casimir scaling}, which has also been observed to hold for the four-loop cusp anomalous dimension~\cite{Moch:2018wjh,Henn:2019swt,vonManteuffel:2020vjv}. 
We stress that we have computed $\gamma_{r,4}^i$ independently for $i\in\{q,g\}$, so that generalized Casimir scaling was not used as an input to our computation.

\section{Energy-Energy Correlation at N$^4$LL}
In this section we use our new result for $\gamma_r^q$ to obtain the first resummation for an event shape at N$^4$LL.
In particular, we consider the Energy-Energy Correlation~\cite{Basham:1978bw} (EEC) in electron-positron annihilation, 

{\footnotesize
\begin{align} \label{eq:EECdef}
    \text{EEC}(\chi) = \sum_{a,b} \int \df \sigma_{e^+ e^-\to a+b+X}\, \frac{E_a E_{b}}{Q^2}\, \delta(\cos\chi_{ab}-\cos\chi) \, , 
\end{align}}%

which was one of the first infrared and collinear safe observables proposed for an $e^+e^-$ collider. 
The EEC measures the angle $\chi_{ab}$ between two final state particles weighted by the energies of the particles relative to the total center-of-mass energy of the colliding $e^+ e^-$ pair. 
Furthermore, the EEC is symmetrized over all possible final state particle pairs, as implemented by the sum in~\eq{EECdef}.
It is convenient to introduce a change of variables and to express the EEC in terms of 
$z \equiv \frac{1}{2} (1 - \cos\chi)$, $z\in[0,1]$.
The small angle limit $(\chi \to 0)$ is reproduced by the $z \to 0$ limit, and the $z \to 1$ limit describes the dijet/back-to-back $(\chi \to \pi)$ configuration. In these limits, the observable becomes strongly sensitive to collinear configurations of the QCD radiation generating large logarithms whose presence spoils the convergence of the perturbative expansion in the strong coupling constant. An all-order understanding in the coupling, which allows for the resummation of these logarithms, can be achieved using factorization theorems \cite{Collins:1981uk,Collins:1981va,Kodaira:1981nh,Kodaira:1982az,Konishi:1978ax,Catani:1992ua,deFlorian:2004mp,Tulipant:2017ybb,Moult:2018jzp,Dixon:2019uzg,Ebert:2020sfi,Vita:2022xxx}.

Throughout its history the EEC has provided the playground for exploring a variety of crucial aspects of QCD and non abelian quantum field theories in general, such as maximally supersymmetric Yang-Mills theory (\SYM). 
As a matter of fact, not only the EEC has been measured in multiple experiments \cite{CELLO:1982rca,JADE:1984taa,Wood:1987uf,TASSO:1987mcs,TOPAZ:1989yod,Abreu:1990us,Acton:1991cu,OPAL:1993pnw,Abreu:1993kj,Abe:1994mf}, but it has been at the intersection of a variety of different theoretical fields. The EEC has been studied at strong coupling using the AdS/CFT correspondence~\cite{Hofman:2008ar}, perturbatively in \SYM \cite{Belitsky:2013xxa,Belitsky:2013bja,Belitsky:2013ofa,Henn:2019gkr,Moult:2019vou,Korchemsky:2019nzm,Kologlu:2019mfz} and in QCD \cite{Kodaira:1982az,Catani:1992ua,deFlorian:2004mp,DelDuca:2016csb,Tulipant:2017ybb,Dixon:2018qgp,Luo:2019nig,Moult:2018jzp,Dixon:2019uzg,Ebert:2020sfi,Gao:2020vyx,Li:2021zcf}, and it constitutes one of the simplest example of energy correlators which have spurred renewed interest in exploring the connections between QCD and $\cN=4$, see for example \cite{Chicherin:2020azt,Henn:2020omi,Chen:2020uvt,Chen:2020adz,Chang:2022ryc,Komiske:2022enw}. Moreover, the EEC can be used for the extraction of the strong coupling constant (see for example \cite{OPAL:1993pnw,Abe:1994mf,Kardos:2018kqj}), and its generalizations to $ep$ and hardon colliders as high precision probe for TMD physics at present and future colliders~\cite{Gao:2019ojf,Li:2020bub,Li:2021txc,Accardi:2012qut,AbdulKhalek:2021gbh,Neill:2022lqx}.

\subsection{EEC in the back-to-back limit}

The back-to-back asymptotics of the EEC can be described using Soft and Collinear Effective Theory (SCET)~\cite{Bauer:2000ew, Bauer:2000yr, Bauer:2001ct, Bauer:2001yt} via the following factorization theorem \cite{Vita:2022xxx}
\begin{align} \label{eq:EEC_fact_thm_q}
 \frac{\df\sigma}{\df z} & = \frac{\born}{8} H_{q\bar q}(Q,\mu) \int_0^\infty \df (b_T Q)^2 \, J_0\bigl(b_T Q \sqrt{1-z}\bigr)
   \\&\quad \times\cJ_q\Bigl(b_T, \mu, \frac{Q b_T}{ \upsilon}\Bigr) \cJ_\bq\Bigl(b_T, \mu, Q b_T \upsilon\Bigr)  [ 1 + \cO(1-z) ] \nn
\,.\end{align}
In \eq{EEC_fact_thm_q}, $J_0$ is the Bessel function arising from the Fourier transform due to the azymuthal symmetry of the EEC measurement, $H_{q\bar q}$ is the quark color singlet SCET hard function, which is related to the IR finite part of the quark form factors \cite{Kramer:1986sg, Matsuura:1987wt, Matsuura:1988sm, Gehrmann:2005pd, Moch:2005tm, Moch:2005id, Baikov:2009bg, Lee:2010cga, Gehrmann:2010ue,Lee:2022nhh} and  can be extracted up to 4 loops from the recent result of \refcite{Lee:2022nhh}, and $\cJ_q$ is the quark EEC jet function which is known up to N$^3$LO \cite{Ebert:2020sfi,Vita:2022xxx}.

The EEC in the back-to-back limit is a \SCETII observable, and therefore requires the handling of rapidity divergences \cite{Collins:1981uk,Ji:2004wu,Beneke:2003pa, Chiu:2007yn, Becher:2011dz,Chiu:2011qc, Chiu:2012ir,Chiu:2009yx, GarciaEchevarria:2011rb,Li:2016axz,Ebert:2018gsn}.
\Eq{EEC_fact_thm_q} is derived in pure rapidity renormalization \cite{Ebert:2018gsn,Vita:2022xxx}, with $\upsilon$ being the pure rapidity renormalization scale. 
In this renormalization scheme, the soft function is dropped since it is 1 to all orders, while the collinear and anti-collinear jet functions are identical up to $\upsilon \to 1/\upsilon$, and the rapidity scale dependence cancels exactly at each order in perturbation theory in the product of the jet functions.

The hard and jet functions in \eq{EEC_fact_thm_q} obey the following renormalization group equations \cite{Vita:2022xxx}, 
\begin{align} \label{eq:RGEs}
 \mu \frac{\df}{\df\mu}{\ln H_{q\bq}(Q,\mu)} &= \gamma^q_H(Q,\mu)\,,
 \nn\\
  \mu \frac{\df}{\df \mu} \ln \cJ_q\Bigl(b_T, \mu, \frac{Q b_T}{ \upsilon}\Bigr) &=  \gamma_{\cJ_q}(\mu,\upsilon \mu/Q)
\,,\end{align}
with the anomalous dimensions
\begin{align} \label{eq:mu_anom_dims}
 \gamma^q_H(Q,\mu) &= 4 \GammaC^{q}[\as (\mu)] \ln\frac{Q}{\mu} + 4 \gamma^q_H[\as(\mu)]
\,,\\
 \gamma_{\cJ_q}(\mu,\upsilon \mu/Q) &= 2 \GammaC^q[\as(\mu)] \ln\frac{\upsilon \mu}{Q} - 2 \gamma^q_H[\as(\mu)]\nn
\,,\end{align}
where $\GammaC^{q}$ is the cusp anomalous dimension in the fundamental representation \cite{Korchemsky:1987wg,Bern:2005iz,Henn:2019swt}, the quark anomalous dimension $\gamma^q_H[\as(\mu)]$ is related to the quark collinear anomalous dimension~\cite{Agarwal:2021zft}, and we differentiated the anomalous dimensions from their non-cusp part by the number of arguments as commonly done in SCET literature.
The EEC jet function also obeys a rapidity RGE, governed by the rapidity anomalous dimension
\begin{align} \label{eq:RRGE}
 \upsilon \frac{\df}{\df \upsilon} \ln \cJ_q\Bigl(b_T, \mu, \frac{Q b_T}{ \upsilon}\Bigr) &= -\frac12 \gamma_r^q(b_T,\mu)
\,,\end{align} 

We solve these RGEs to obtain the resummed cross section for the EEC explicitly in terms of the anomalous dimensions and boundary functions
\begin{align} \label{eq:EEC_resummed}
 \frac{\df\sigma}{\df z} &
 = \frac{\born}{8} \int_0^\infty \!\!\df (b_T Q)^2 \, J_0\bigl(b_T Q \sqrt{1-z}\bigr)H_{q\bq}(Q,\mu_H)
 \\ \times\,&
 \cJ_q\Bigl(b_T, \mu_J, \frac{Q b_T}{ \upsilon_n}\Bigr) \cJ_\bq\Bigl(b_T, \mu_J, Q b_T \upsilon_\bn \Bigr)  \Bigl(\frac{\upsilon_n}{\upsilon_\bn}\Bigr)^{ \frac12\gamma_r^q(b_T, \mu_J)}
 \nn\\ \times\,&
 \exp\left[ 4\int_{\mu_J}^{\mu_H} \frac{\df\mu'}{\mu'} \GammaC^q[\as(\mu^\prime)] \ln\frac{\mu^\prime}{Q} -  \gamma^q_H[\as(\mu^\prime)]\right]
\nn\,.\end{align}

{
\renewcommand{\arraystretch}{1.2}
\begin{table}[pt]
\centering
 \begin{tabular}{l|c|c|c|c|c} \hline\hline
  Accuracy & $H$, $\cJ$ & $\GammaC(\as)$  & $\gamma^q_H(\as)$ & $\gamma^q_r(\as)$ & $\beta(\as)$ \\\hline
  LL           & Tree level & $1$-loop & --       & --       & $1$-loop \\\hline
  NLL          & Tree level & $2$-loop & $1$-loop & $1$-loop & $2$-loop \\\hline
  NLL$^\prime$ & $1$-loop   & $2$-loop & $1$-loop & $1$-loop & $2$-loop \\\hline
  NNLL         & $1$-loop   & $3$-loop & $2$-loop & $2$-loop & $3$-loop \\\hline
  NNLL$^\prime$& $2$-loop   & $3$-loop & $2$-loop & $2$-loop & $3$-loop \\\hline
  N$^3$LL         & $2$-loop   & $4$-loop & $3$-loop & $3$-loop & $4$-loop \\\hline
  N$^3$LL$^\prime$& $3$-loop   & $4$-loop & $3$-loop & $3$-loop & $4$-loop \\\hline
  N$^4$LL         & $3$-loop   & $5$-loop & $4$-loop & $4$-loop & $5$-loop \\\hline
  N$^4$LL$^\prime$& $4$-loop   & $5$-loop & $4$-loop & $4$-loop & $5$-loop \\\hline
 \hline
 \end{tabular}
\caption{%
Resummation accuracy in terms of the perturbative order of boundary terms, anomalous dimensions and beta function.
}
\label{tbl:log_counting}
\end{table}
\renewcommand{\arraystretch}{1.0}
}

The logarithmic accuracy of the resummed cross section is defined in terms of the perturbative order at which the ingredients entering \eq{EEC_resummed} are computed, as shown in Table \ref{tbl:log_counting}. 
Explicitly, N$^4$LL resummation requires the cusp anomalous dimension and the QCD beta function to 5 loops \cite{Herzog:2018kwj,Baikov:2016tgj}, the collinear dimension at 4 loops \cite{Agarwal:2021zft}, the jet function boundaries at 3 loops \cite{Ebert:2020sfi}, the hard function at 3 loops \cite{Lee:2010cga,Gehrmann:2010ue}, and the 4-loop rapidity anomalous dimension, which we obtained in this Letter.
In combination with an approximation of the five loop cusp anomalous dimension~\cite{Herzog:2018kwj}, we have now all anomalous dimensions for N$^4$LL resummation at our disposal and can apply them towards realistic observables.

\subsection{Numerical Results}
\begin{figure}[h]
 \centering
  \includegraphics[width=0.48\textwidth]{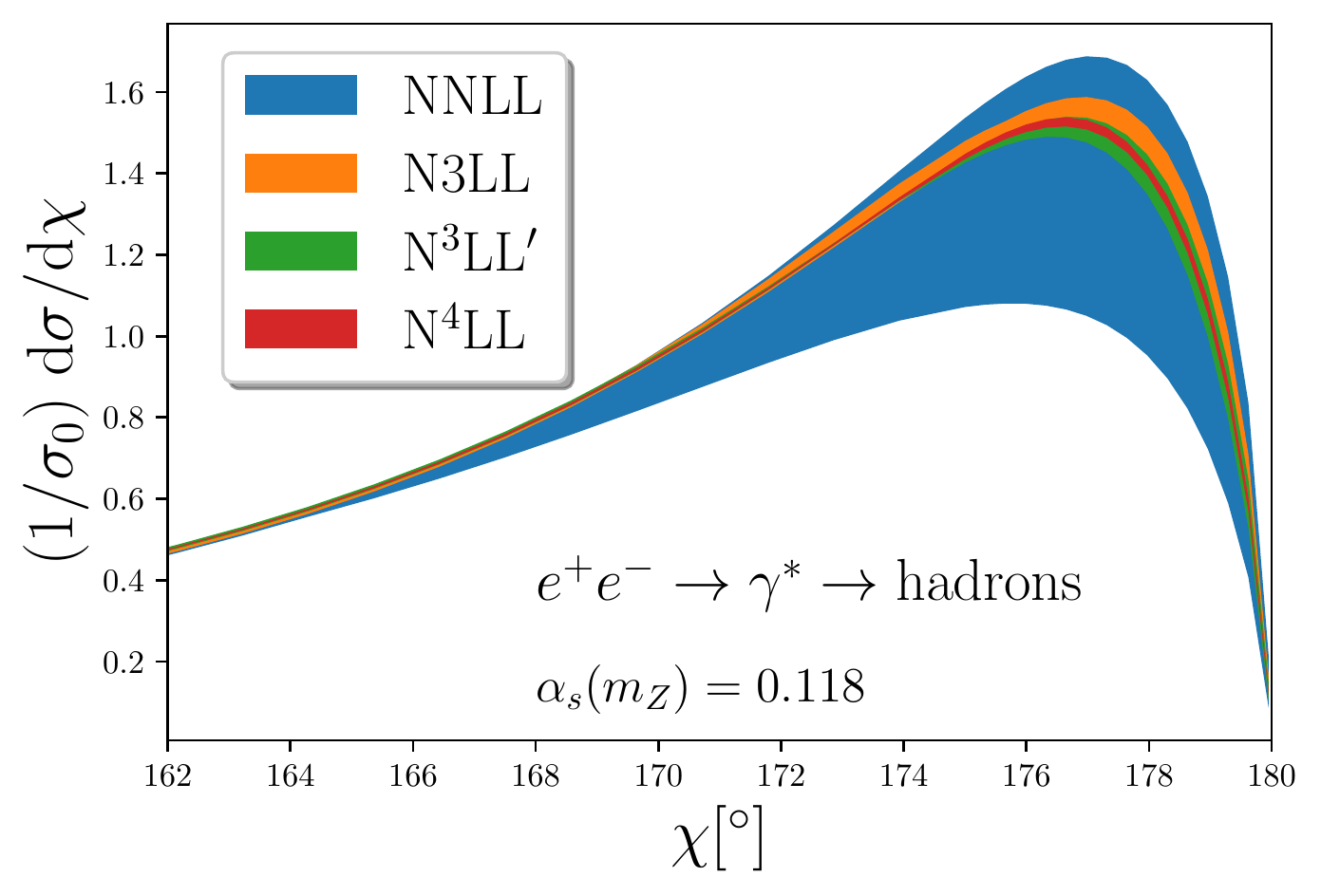}
 \caption{Resummed result for the EEC in the back-to-back region up to N$^4$LL accuracy. Uncertainty bands reflect the residual perturbative uncertainty and are obtained with a 15-point scale variation of the resummation scales. See text for details.}
 \label{fig:EEC_plot}
\end{figure}
We have implemented the resummed cross section of \eq{EEC_resummed} in a private python code and performed the resummation of this observable up to N$^4$LL.
Note that this constitutes the first ever resummation for an event shape at this level of accuracy. 
On top of all the necessary ingredients for N$^4$LL resummation, we also include the 4-loop hard function, which we have extracted from the 4-loop form factor calculation of \refcite{Lee:2022nhh}.
Figure~\ref{fig:EEC_plot} shows our results as a function of the scattering angle $\chi$ through different logarithmic orders. 
We observe that increasing the logarithmic order leads to an improved description of the EEC.
We indicate uncertainty estimates due to the truncation of the logarithmic accuracy by colored bands and observe that successively higher order bands are contained within the estimates based on previous orders. 
We conclude that the our computation of the EEC in the limit of $z\to 1$ at N$^4$LL yields a highly precise determination of the perturbative contribution to scattering observable in this limit.

Our uncertainty estimates are based on the variation of renormalization scales.
As expected, the explicit dependence on the renormalization scales $\mu$ and $\upsilon$ exactly cancels in the resummed cross section in~\eq{EEC_resummed}. 
The result depends on the boundary scales $\{\mu_H,\mu_J,\upsilon_{n,\bn}\}$ marking the starting points of the RG evolution. 
The choice of these boundary scales is in principle arbitrary and, at any given logarithmic accuracy, the resummed cross sections obtained with different choices of boundary scales would give results that differ by terms that are beyond this logarithmic accuracy.
We select the scales: 
\beq
	\{\mu^{*}_H = Q,\mu^{*}_J = b_0/b_T,\upsilon^{*}_{n} = Q b_T/b_0 = 1/\upsilon^{*}_{\bn}\}.
\eeq
When choosing these values for the boundary scales, all explicit logarithms in the boundary functions vanish identically.
\Eq{EEC_resummed} evaluated with this canonical choice constitutes our central value of the resummed prediction.
We estimate perturbative uncertainties on the resummed cross section by evaluating \eq{EEC_resummed} with different boundary scales. Here, we vary the scales individually by a factor of $\frac{1}{2}$ or 2 around their canonical value and remove the configurations with simultaneous variations of factors greater than 2 or smaller than $\frac{1}{2}$.
Next, we take the envelope of the results as our estimate of the perturbative uncertainty.
This results in a 15-point scale variation procedure very analogous to the usual 7-point scale variation employed to estimate perturbative uncertainties in fixed order calculations.
To treat the large $b_T$ behavior in the Fourier transform we use the $b^*$ prescription \cite{Collins:1981uk,Collins:1981va} employed in \refcite{Ebert:2020sfi}.

\begin{figure}
 \centering
  \includegraphics[width=0.48\textwidth]{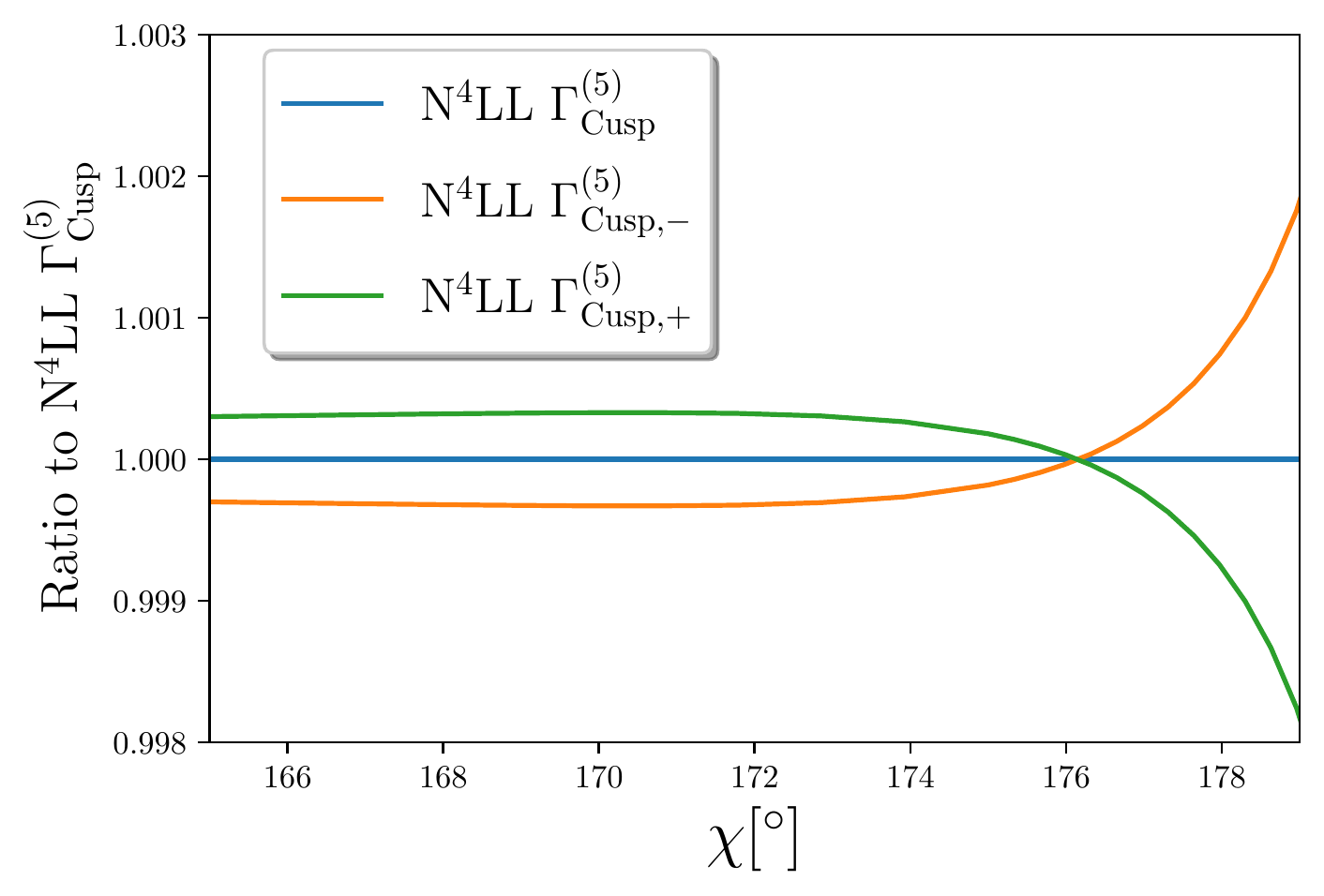}
 \caption{Comparison of the central value for the EEC distribution between the resummed result computed with different values of the 5-loop cusp anomalous dimension.}
 \label{fig:cusp_impact}
\end{figure}

Note that the cusp anomalous dimension is known at 5 loops only in approximate form \cite{Herzog:2018kwj} with an 80\% relative uncertainty, $\GammaC^{(5)}= 0.21 \pm 0.17$, but it is in general expected that its numerical impact to be very small. In \fig{cusp_impact} we show the effect of varying the 5 loops cusp anomalous dimension coefficient around the values of the uncertainty, $\{\Gamma_{\rm Cusp,+}^{(5)}= 0.38,\Gamma_{\rm Cusp}^{(5)}= 0.21,\Gamma_{\rm Cusp,-}^{(5)}= 0.04\}$. 
We see that it generates a sub-per-mille variation, confirming that it is indeed the case that its numerical impact is small and that the approximation of \refcite{Herzog:2018kwj} is more than enough for current phenomenological studies.

We leave a full phenomenological study of the EEC including fixed order predictions~\cite{Dixon:2018qgp,DelDuca:2016csb,Tulipant:2017ybb}, state of the art resummation in the $z \to0$ limit~\cite{Konishi:1978ax,Dixon:2019uzg} as well as estimation of parametric and non-perturbative uncertainties to future work.

\section{Conclusion}
Throughout this Letter we have discussed the computation of the four-loop corrections to the quark and gluon rapidity anomalous dimensions, which control the all-order structure of large logarithms for several quantities of phenomenological interest, including transverse momentum distributions at proton colliders and event shape observables at $e^+e^-$ colliders.
Our computation is built on our recent determination of the four-loop soft anomalous dimension and the conjectured duality between the soft and rapidity anomalous dimensions. 
Our result is fully analytic, up to four constant that are only known numerically. 
Remarkably, our results exhibit generalized Casimir scaling, a property which was observed to hold also for the cusp anomalous dimension through four loops. 
We also applied our results for the rapidity anomalous dimension to obtain for the first time phenomenological results for the EEC in the back-to-back region at N$^4$LL, providing the most precise resummed calculation for this observable to date and the first example of the resummation of a TMD observable to fourth logarithmic order. 
This shows that our result will play an important role in the future precisely determine several quantities of phenomenological interest.

\begin{acknowledgments}
\emph{Acknowledgements:}  
We thank Ian Moult, HuaXing Zhu, and YuJiao Zhu for coordinating the submission of their work \cite{Moult:2022xzt}, and YuJiao Zhu for pointing out a typo in eq.~\eqref{eq:bdef} in the first version of the preprint.
GV thanks Lance Dixon and Duff Neill for useful discussions. 
GV and BM are supported by the United States Department of Energy, Contract DE-AC02-76SF00515. 
\end{acknowledgments}

\bibliography{refs}

\end{document}